\documentclass[compsoc, conference, a4paper, 10pt, times]{IEEEtran}

\usepackage{cite}
\usepackage{amsmath,amssymb,amsfonts}
\usepackage{algorithmic}
\usepackage{graphicx}
\usepackage{textcomp}
\usepackage{xcolor}
\usepackage{booktabs}
\usepackage[hidelinks]{hyperref}
\usepackage{multicol}
\usepackage{colortbl}

\begin{document}

\title{A Qualitative Analysis Framework for mHealth Privacy Practices}


\iftrue
\author{\IEEEauthorblockN{Thomas Cory}
\IEEEauthorblockA{\textit{Technische Universität Berlin} \\
Berlin, Germany \\
cory@tu-berlin.de}
\and
\IEEEauthorblockN{Wolf Rieder}
\IEEEauthorblockA{\textit{Technische Universität Berlin} \\
Berlin, Germany \\
w.rieder@tu-berlin.de}
\and
\IEEEauthorblockN{Thu-My Huynh}
\IEEEauthorblockA{\textit{Technische Universität Berlin} \\
Berlin, Germany \\
t.huynh.1@tu-berlin.de}
}
\fi

\maketitle

\begin{abstract}

Mobile Health (mHealth) applications have become a crucial part of health monitoring and management. However, the proliferation of these applications has also raised concerns over the privacy and security of Personally Identifiable Information and Protected Health Information. Addressing these concerns, this paper introduces a novel framework for the qualitative evaluation of privacy practices in mHealth apps, particularly focusing on the handling and transmission of sensitive user data. Our investigation encompasses an analysis of 152 leading mHealth apps on the Android platform, leveraging the proposed framework to provide a multifaceted view of their data processing activities.

Despite stringent regulations like the General Data Protection Regulation in the European Union and the Health Insurance Portability and Accountability Act in the United States, our findings indicate persistent issues with negligence and misuse of sensitive user information. We uncover significant instances of health information leakage to third-party trackers and a widespread neglect of privacy-by-design and transparency principles. Our research underscores the critical need for stricter enforcement of data protection laws and sets a foundation for future efforts aimed at enhancing user privacy within the mHealth ecosystem.
\end{abstract}

\section{Introduction}
\label{sec:intro}

Mobile Health (mHealth) applications have become increasingly popular in recent years, with many users relying on these apps to track and monitor their health data~\cite{gminsights2022mhealth}. However, the widespread use of these apps has raised concerns about the privacy and security of personal health information. In particular, the collection and use of Personally Identifiable Information (PII) and Protected Health Information (PHI) by mHealth apps has become a topic of growing concern for individuals, regulators, and policymakers alike~\cite{scatterday2021}.

The regulation of personal information and health data varies across regions and jurisdictions, with the European Union and the United States implementing laws to protect individuals' privacy rights. In the European Union, the General Data Protection Regulation (GDPR)~\cite{gdpr} requires that health data be processed only for specific purposes and with explicit consent from the data subject and mandates that appropriate technical and organisational measures be taken to ensure the security and transparency of such data processing. In the United States, the Health Insurance Portability and Accountability Act (HIPAA)~\cite{hipaa} and the Federal Trade Commission (FTC) Act~\cite{ftc} establish similar legal requirements.

Despite these regulations, tracking and advertising appear widespread in the mHealth industry~\cite{bloomberg}, with alarming instances of data privacy breaches gaining public attention in recent years. Notably, the FTC raised allegations of privacy breaches against Flo~\cite{ftcvflo} and Easy Healthcare Corporation~\cite{ftcvprenom}, two prominent developers of menstrual tracking apps. Additionally, the recent repeal of Roe v. Wade~\cite{roevwade}, a decision by the United States Supreme Court that reversed a key ruling on abortion rights~\cite{ziegler}, has raised concerns that health data collected by mHealth apps, particularly those related to menstrual tracking, could be misused in ways that infringe on personal privacy rights~\cite{nprroevwade}.

In this study, we tackle the pivotal question of how mHealth apps manage users' sensitive PII and PHI against the backdrop of evolving privacy regulations and increasing concerns over data sovereignty. Specifically, our contributions are as follows:

\begin{enumerate}
    \item We introduce a comprehensive framework that enables a qualitative evaluation of mHealth data practices by enriching the analysis of PII and PHI transmissions with four supplementary data sources, offering a multi-dimensional view of mHealth privacy practices that extends beyond conventional quantitative approaches.

    \item Through an in-depth examination of 152 leading Android mHealth apps, we utilise our novel qualitative analysis framework to uncover nuanced insights into data management practices of mHealth apps, including compliance with privacy regulations and alignment with user expectations.

    \item Lastly, we critique the effectiveness and accuracy of privacy labels in reflecting actual data practices, revealing discrepancies between declared and real-world app behaviours.
\end{enumerate}

These contributions provide critical insights into the privacy dynamics of mHealth applications, addressing gaps in the current understanding of data management practices and their implications for user privacy and data protection.

\section{Related Work}
\label{sec:rw}

The technological capabilities of mHealth applications and their processing of health information entail a series of privacy challenges that researchers still work to address to this day. On the predictive side, this area of research covers taxonomies~\cite{Ozdalga2012TheSI}~\cite{plachkinova} and frameworks~\cite{avancha}~\cite{stavrou}, the analysis of privacy policies~\cite{Sunyaev2015AvailabilityAQ}, as well as user surveys and interviews~\cite{schroeder}~\cite{Atienza}. These research avenues are essential to establish baselines for evaluating the privacy-related behaviour of mHealth apps but fall short of capturing their actual behaviour. This gap is addressed by several studies that follow similar lines of research to this paper by performing empirical analyses of the data processing practices of mHealth apps.

In 2015, Knorr et al.~\cite{knorr} examined 154 Android hypertension and diabetes mHealth apps based on an analysis of their privacy policies, supported by a technical analysis of their data handling practices. The latter combined static and dynamic code analysis approaches, as well as a network traffic analysis for apps that communicated with remote servers. This analysis was expanded upon by Huckvale et al.~\cite{Huckvale2015}, who put forward two key findings: firstly, tracking is rarely a one-off event, and secondly, inconsistencies between information provided in privacy policies and the actual behaviour of the observed mHealth apps are not uncommon.

More recently, in 2018, Papageorgiou et al.~\cite{Papageorgiou} examined 20 Android mHealth apps using a combination of automated static code and traffic analysis. They manually interacted with each application to simulate real user data, finding widespread transmission of health-related data to third parties. Similarly, Grundy et al. \cite{grundy} analysed the transmission of user and device data to third parties based on manual interactions with a set of 24 medical apps. Their traffic analysis revealed that 79\% of apps transmitted user data ranging in sensitivity from device identifiers to PHI.

Taking a different approach to the manual execution and analysis of small sets of mHealth apps, Tangari et al.~\cite{Tangari} performed an automated traffic and code analysis on 15\,893 free apps, with traffic data collected via an external Man-in-the-Middle proxy. PII transmission and leakage were identified by applying Ren et al.'s machine learning approach~\cite{recon}, with results showing that the transmission of personal data was lower for \textit{Medical} apps than for \textit{Health and Fitness} apps. Notably, Tangari et al. only observed transmissions of user data in a comparatively small subset of 616 apps, which can be traced back to the lack of user interaction and data entered during the automated crawl.

Similar approaches have also been applied to specific mHealth domains, such as diabetes~\cite{blenner}, dementia~\cite{rosenfeld}, depression~\cite{depression} and female health~\cite{malki2024exploring}.

Although these studies provide detailed insights into individual dimensions of mHealth privacy practices, they generally fall short of integrating multiple perspectives into a holistic view that enables a qualitative assessment of observed mHealth privacy practices. 

\section{Concept}
\label{sec:meth}

In the complex and rapidly evolving domain of mHealth applications, safeguarding sensitive user information, particularly health-related information, is paramount. Understanding how mHealth apps manage these sensitive data types is crucial to assessing their privacy practices. At the heart of our investigation is the question:

\textit{How do mHealth apps manage the sensitive PII and PHI of their users?}

This question sets the stage for a comprehensive investigation of the operational, ethical, and legal dimensions of data management in mHealth applications. Simple quantitative assessments are insufficient to provide the context necessary to understand the complex dynamics of mHealth data practices. Addressing this shortcoming necessitates a methodology that can differentiate privacy practices on a spectrum ranging from compliance to non-compliance, from beneficial to detrimental, and from alignment to misalignment with user expectations and regulatory standards.

\begin{figure}[tb]
  \centering
  \includegraphics[width=\linewidth]{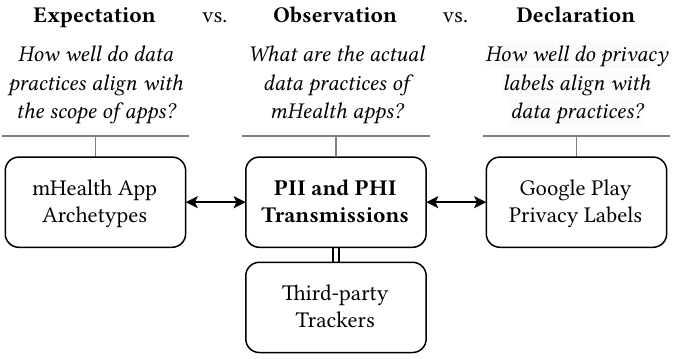}
  \caption{Overview of the proposed analysis framework, highlighting the three integrated dimensions that provide the context for a qualitative assessment of mHealth privacy practices.}
  \label{fig:approach}
\end{figure}

We propose a methodological framework that adopts a qualitative approach to evaluating the privacy practices of mHealth applications. This approach complements and expands upon quantitative assessments by contrasting actual data management practices with expected and declared practices. As depicted in Figure~\ref{fig:approach}, our framework synthesises the dimensions of \textit{Observation}, \textit{Expectation}, and \textit{Declaration} to furnish a comprehensive perspective on the privacy practices of mHealth apps. By addressing three supplemental guiding questions along these dimensions, this triangulation enables a nuanced evaluation of the alignment between app practices, user expectations, and legal standards.

\subsection{Observation}
\label{subsec:meth:observation}

Central to our framework is a detailed quantitative analysis that addresses the guiding question:

\textit{What are the actual data practices of mHealth applications?}

We focus this quantitative analysis on the following key aspects of mHealth privacy practices: 

\begin{enumerate}
    \item \textbf{Embedded Trackers}: An investigation into the use of third-party trackers' APIs and libraries embedded within the code of mHealth applications assesses the prevalence of third-party tracking, the potential PII and PHI available to these trackers, and resulting privacy implications.
    \item \textbf{Contacted Trackers}: An analysis of the network traffic of mHealth apps reveals transmission streams to third-party trackers and identifies patterns of data sharing and leakage, complementing the insights gained from the investigation of embedded trackers.
    \item \textbf{PII and PHI Transmissions}: Building upon the insights gained from investigating the prior aspects, a detailed examination of PII and PHI transmissions within the HTTP traffic generated by mHealth apps uncovers data management practices and specific instances of sensitive data leakage.
\end{enumerate}

By integrating these aspects, the quantitative analysis forms the basis for a more comprehensive qualitative assessment by providing observations of actual mHealth privacy practices to be contextualised through a juxtaposition with the other two dimensions of our framework.

\subsection{Expectation}
\label{subsec:meth:expectation}

The GDPR lays out seven fundamental principles for the processing of personal data, chief among which are \textit{data minimisation} and \textit{purpose limitation}. 
Put together, these two principles require that the scope and purpose of data processing by apps be limited to what is strictly necessary for their functionality. We assess the adherence of mHealth apps to these principles by addressing the guiding question:

\textit{How well do mHealth apps' data practices align with their scope and purpose?}

In order to derive the expected scope of PHI available to mHealth apps, we cross-reference their primary features with the framework of mHealth app archetypes proposed by Dehling et al. \cite{dehling2015exploring}. This framework identifies three distinct levels of PHI specificity that mHealth apps are presumed to access: \textit{standard} information encompasses all non-health-related PII, \textit{nonstandard} information refers to non-medical PHI (e.g. daily step count), and \textit{medical} information represents the highest level of PHI that is only accessible by dedicated medical applications (e.g. medical records).

This baseline of expected behaviour provides the context required to evaluate whether mHealth apps are compliant with the principles of data minimisation and purpose limitation by assessing whether their data transmissions lie within the scope of data expressly required for their purpose.

\subsection{Declaration}
\label{subsec:meth:declaration}

Another core principle of the GDPR is \textit{transparency} as the main enabler of informed consent, which is an important basis for the lawful collection of personal data. This principle requires the disclosure of data management practices prior to the collection of personal data in order to provide data subjects with the insights necessary to exercise their data rights. 

In the mobile domain, \textit{privacy labels} have emerged as a prominent method for app developers to provide a succinct overview of their apps' data practices. As highlighted by Rodriguez et al.~\cite{rodriguez2023labels}, however, the quality of these self-declared privacy labels varies between ecosystems and app categories.

By assessing the overlap between privacy labels of mHealth apps and their observed data transmissions, we evaluate the transparency and accuracy of privacy labels in reflecting actual data practices of mHealth apps, thereby addressing the guiding question:

\textit{How well do privacy labels reflect the actual data practices of mHealth apps?}

\section{Implementation}
\label{sec:impl}

We evaluate the proposed framework by applying it to the most popular mHealth apps available on Google Play, the primary distribution service for Android apps. To this end, we generate a dataset of network traffic by manually interacting with the apps, following the hypothesis that automated tools are unable to accurately emulate real user behaviour due to the high level of reasoning required to interact with the vast variety of app features and UI design patterns present in modern mHealth apps.

\subsection{App Selection}
\label{subsec:impl:appsel}

Collecting the data necessary for such an investigation requires a high degree of manual input and meticulous data entry to maintain accuracy and integrity in the dataset. This limits the number of apps that can feasibly be examined compared to fully automated crawls. Therefore, we restrict our selection to the most popular apps (based on their total number of reviews and downloads) from the relevant Google Play categories \textit{Health and Fitness} and \textit{Medical}, and apply a set of filters to ensure that all selected apps are fully functional and accessible during our interactions with them. Specifically, we exclude all apps that do not fulfil the following requirements:

\begin{enumerate}
    \item App can be downloaded and executed without restrictions in Germany.
    \item App is executable on a rooted device.
    \item Primary app features do not employ measures to prevent the decryption of HTTPS traffic.
    \item Primary app features are accessible for free.
    \item No official identification (e.g. health insurance number) is required to access the primary features.
\end{enumerate}

Applying this filter to the 500 most popular apps from the relevant Google Play categories returns 152 mHealth apps that are eligible for our analysis.

\subsection{Data Collection}
\label{subsec:impl:datacol}


We generate the primary dataset for our analysis by executing these 152 mHealth apps on a Google Pixel 6a running a rooted version of Android 13. To emulate real user interactions, we manually interact with each app in our selection while following a standardised interaction pattern, which is defined as follows: 

\begin{enumerate}
    \item Create user accounts and log into apps whenever possible to ensure consistent access to primary app features.

    \item Interact with all feasibly reachable primary features.

    \item Enter personal and health-related information wherever applicable to inject as much traceable PII and PHI as possible.

    \item Agree to all privacy-related prompts to maximise the potential for observable data transmissions.
\end{enumerate}

\begin{figure}[tb]
  \centering
  \includegraphics[width=\linewidth]{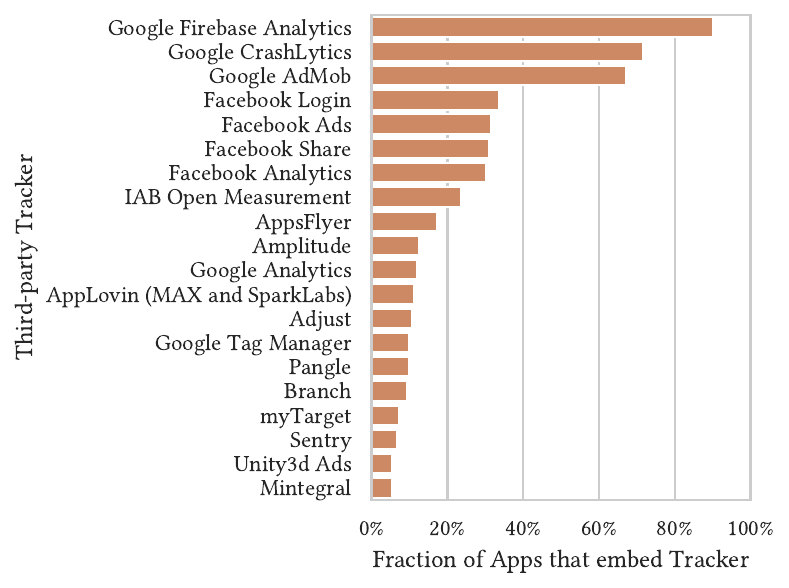}
  \caption{The 20 most prevalent third-party trackers ranked by the number of mHealth apps that embed their Java libraries.}
  \label{fig:libraries}
\end{figure}

Throughout the manual execution of the apps, we apply a synthetic persona that covers all types of requested PII and PHI, thereby providing a standardised basis of known values for the detection of PII and PHI leakage.
Since the content format of HTTP messages is not standardised and varies greatly between applications, we use the attributes of this persona as the basis for a thorough keyword-based detection of PII and PHI transmissions in the HTTP traffic generated by the examined apps. Therefore, the persona is expanded iteratively whenever an app requests previously undefined PII or PHI to ensure that it comprises all values of divulged information.

In addition to this manual crawl, we perform an automated crawl in which each app is executed for 60 seconds without user interactions. The network traffic resulting from this non-interactive crawl is used to compare data transmissions caused by manual user interactions and automated crawls and determine whether such automated data collection methods are a suitable replacement for our labour-intensive manual approach.

During both crawls, we employ the \textit{Heimdall} privacy monitoring toolkit~\footnote{https://github.com/tomcory/Heimdall} presented by Cory et al.~\cite{cory2020heimdall} to intercept and record all HTTP messages transmitted during the execution of the monitored mHealth apps. Remote hosts are labelled by \textit{Heimdall} in real time as \textit{trackers} and \textit{non-trackers} based on the \textit{Unified Hosts} file curated by Steven Black~\cite{StevenBlack2023}.

Differentiating data transmissions to first- and third-party hosts is generally difficult in the mobile domain, as first-party hostnames rarely match available app identifiers. As our analysis requires the identification of third parties to discern instances of data sharing, we equate trackers to third parties. Therefore, any data transmission to a tracker host represents an instance of data sharing within the scope of our analysis.

\section{Results}
\label{sec:res}

Following the approach laid out by our proposed qualitative evaluation framework, we divide our analysis along three dimensions. First, we examine the actual privacy practices of mHealth apps through a quantitative analysis of their source code and network traffic. We then contextualise the resulting observations by contrasting observed transmissions of PII and PHI with the expected scope of the examined applications and their data practices as declared by the privacy labels available on Google Play.

\subsection{Observation}
\label{subsec:meth:appsel}

We further divide our quantitative analysis along the three aspects defined by our framework. First, we examine the extent to which the examined apps embed third-party trackers in their source code. We then expand upon this examination by scrutinising the observed network traffic to reveal communications with third-party trackers. This establishes the basis for a thorough review of PII and PHI transmissions in the observed network traffic, in which we identify instances where data is collected and shared with third parties.

\subsubsection{Embedded Trackers}

\begin{figure}[tb]
  \centering
  \includegraphics[width=0.895\linewidth]{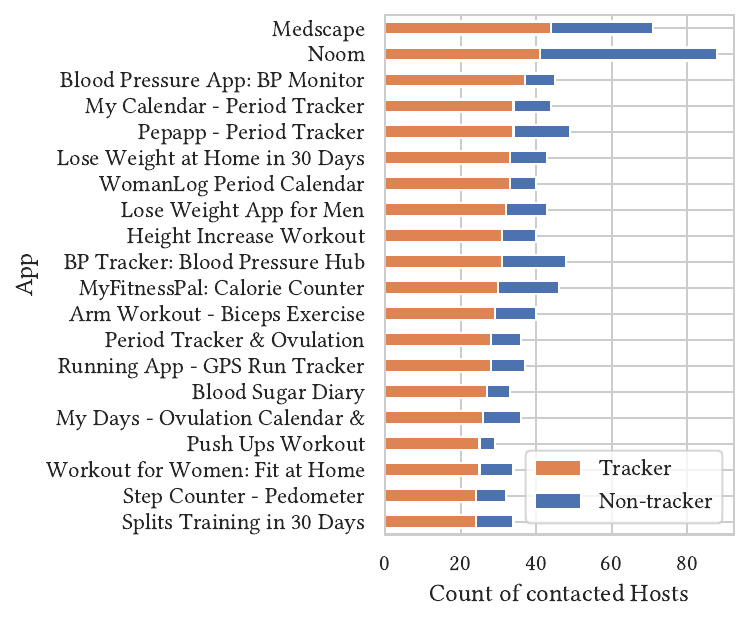}
  \caption{The 20 apps that contacted the highest number of third-party tracker hosts, contrasted with the number of non-tracker hosts they contacted.}
  \label{fig:contacted_hosts}
\end{figure}

\begin{figure*}[ht]
  \centering
  \includegraphics[width=\linewidth]{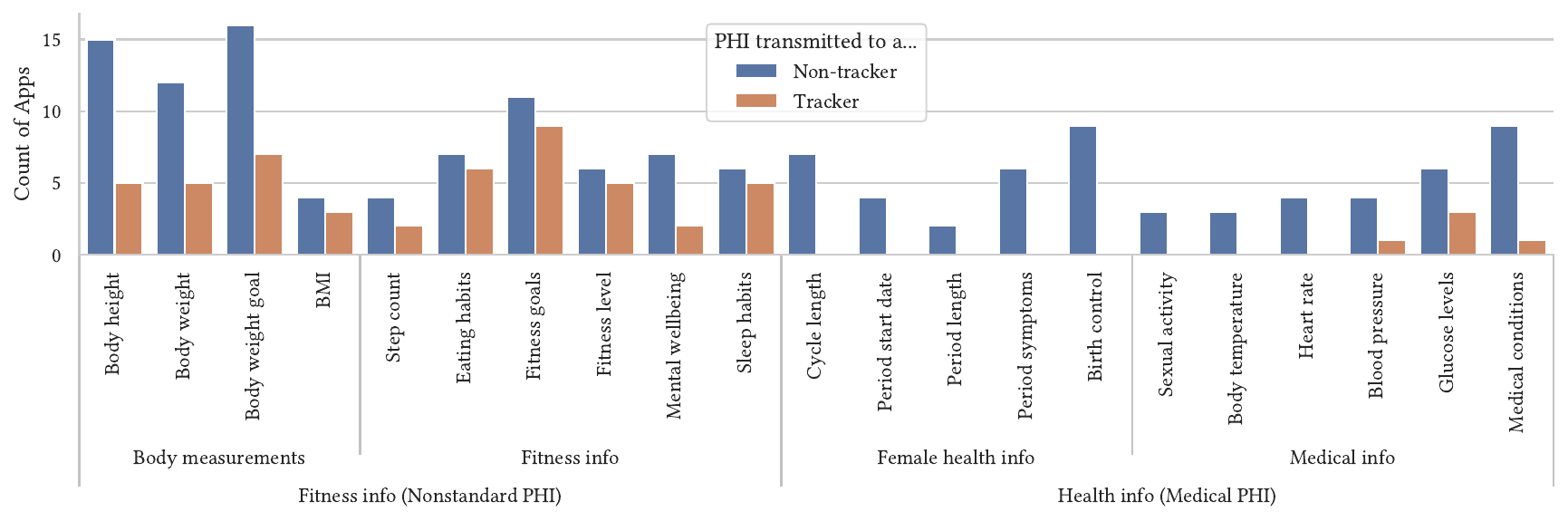}
  \caption{Transmission volume (i.e. number of transmitting apps) of 21 PHI types, grouped by data category and specificity, contrasting transmissions to non-trackers and third-party trackers.}
  \label{fig:transmission_apps_data_types}
\end{figure*}

\begin{figure}[ht]
  \centering
  \includegraphics[width=\linewidth]{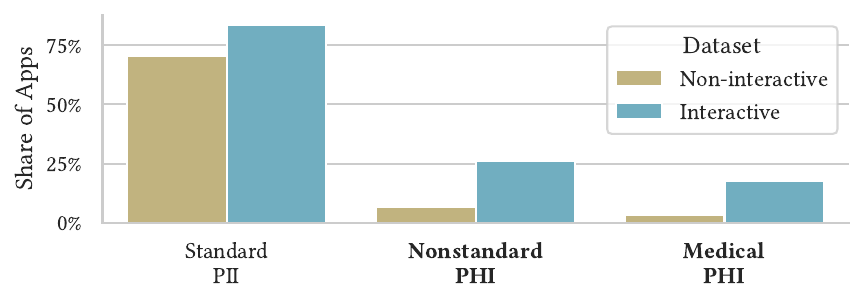}
  \caption{Comparison of observed transmission volumes of standard PII, nonstandard PHI, and medical PHI in the network traffic captured during the manual interactive crawl and the automated non-interactive crawl.}
  \label{fig:transmission_apps_specificities}
\end{figure}

Third-party tracking providers commonly offer convenient libraries and APIs to simplify the integration of their services within applications. These embedded third-party libraries pose substantial privacy concerns as they inherently have the same level of access to user data and device functionality as the apps they are incorporated into.

Analysing the package structure of the studied apps and comparing their class definitions against an extensive database of over 400 tracker libraries curated and maintained by the Exodus Privacy Project~\cite{exodus} reveals that 94.7\% of apps embed at least one third-party tracker, with an average of 6.3 embedded trackers per app. The highest number of trackers is embedded in a \textit{Female Health} app designed to track the menstrual cycle of its users: \textit{Peppapp - Period Tracker} embeds 19 tracker libraries from 14 distinct third-party entities.

Overall, we identify 83 third-party tracking libraries, of which 55 are embedded in at least two distinct apps. Ranking the libraries by the number of apps that embed them, as shown in Figure~\ref{fig:libraries}, highlights a pronounced long-tail distribution in the prevalence of third-party tracking libraries embedded in our sample of mHealth apps: 
the most common libraries are embedded by a majority of apps, with the most common library, \textit{Google Firebase Analytics}, appearing in 90.1\% of apps, whereas the majority of trackers are included in a much smaller number of apps. 
Google and Facebook dominate this distribution, publishing seven of the ten most commonly embedded third-party tracker libraries.

\subsubsection{Contacted Trackers}

Having revealed the widespread use of embedded third-party trackers within the examined mHealth apps, we proceed with an investigation of their prevalence in the apps' network traffic. Overall, during our interactions with them, the examined apps contacted 1,181 unique hostnames and IP addresses across 393 distinct domains, 133 of which are associated with third-party trackers.

Mirroring our observations of the embedded trackers, Google dominates this space, operating nine of the ten most-contacted domains. The most prominent domain, \textit{crashlytics.com}, is contacted by 69\% of the examined apps, followed by \textit{gstatic.com} (57.8\%) and \textit{doubleclick.net} (55.2\%).

The distribution of tracker and non-tracker hosts contacted by the apps is depicted in Figure~\ref{fig:contacted_hosts}, highlighting the apps that communicated with the highest number of tracker hosts. Notably, \textit{Medscape} emerges as the top communicator, interacting with 44 tracker hosts across 23 different domains, closely followed by \textit{Noom}, which communicates with 41 different tracker hosts from 28 domains. Overall, we observe that 94 out of 152 apps contact more tracker hosts than non-tracker hosts, as opposed to just four apps that contact no third-party trackers.

\subsubsection{PII and PHI Transmissions}

The primary goal of our quantitative analysis of the network traffic of the examined mHealth apps is to pinpoint transmissions of PII and PHI. We focus this examination on outbound HTTP requests that feature a non-zero content length, culminating in the review of 7,178 out of a total of 26,519 HTTP requests collected during our interactions with the examined mHealth apps. PII and PHI transmissions in these HTTP requests are identified through a regex-based keyword search encompassing 14 types of non-health-related PII as well as 21 types of PHI. These range from generic fitness metrics, such as body height or eating habits, to more sensitive health data like blood pressure or specific medical conditions.

Our findings reveal a significant disparity in data transmission practices among the apps studied. A substantial 80.4\% of the examined apps are found to transmit non-health-related PII, whereas fitness-related information is transmitted by 19.6\%, and health information by 9.3\% of the apps.

Focusing on transmitted PHI, we observe a trend of decreasing transmission volume with increasing specificity. This is illustrated in Figure~\ref{fig:transmission_apps_data_types}, which depicts the number of apps transmitting each type of PHI to non-trackers and trackers. In general, nonstandard PHI is transmitted more commonly than medical PHI. Body metrics, i.e. body weight and height, emerge as the most commonly transmitted type of PHI, followed by fitness goals and eating habits.

Overall, we observe a notable contrast between the transmission of nonstandard PHI and more sensitive medical PHI. While nonstandard PHI is commonly shared with third parties, medical PHI is less frequently transmitted and seldom observed to be sent to third-party trackers. In particular, we observe no direct transmissions of female health information to third-party trackers. We do, however, observe transmissions of other medical PHI to trackers, specifically blood pressure data, blood glucose levels, and medical conditions such as headaches or even diabetes. The latter is especially concerning, as medical conditions represent the highest specificity of PHI according to the classification of Dehling et al., with the highest potential damage in case of misuse.

\subsubsection{Manual Interactions vs. Automated Crawls}

As highlighted in Figure~\ref{fig:transmission_apps_specificities}, we do not observe the same volume of PHI transmissions in the network traffic stemming from the automated, non-interactive crawl. Although a similar percentage of apps transmitted some form of non-health-related (i.e. standard) PII, the transmission volume of nonstandard and medical PHI is greatly reduced compared to the manual crawl. This confirms our hypothesis that manual interactions are required to generate the network traffic required for this analysis, as the more sensitive PHI is generally not available to apps without further user input.

\subsection{Expectation vs. Observation}
\label{subsec:res:expectation}

While manually interacting with the examined mHealth apps, we identified 14 distinct categories of primary app features. These range from more casual features like \textit{Screen Overlay} apps, which dim the phone's screen to reduce eyestrain, to critical medical apps like \textit{Telemedicine} apps that enable direct interactions between patients and physicians.

By mapping these feature categories to Dehling et al.'s app archetype framework as described in Section~\ref{subsec:meth:expectation}, we establish a baseline for the highest expected PII and PHI specificity available to each app. We identify two casual feature categories that should only access standard PII, six categories limited to nonstandard PHI and six categories whose scope encompasses highly sensitive medical PHI.

Mapping all types of PII and PHI observed during the examination of the apps' network traffic to these specificity levels enables us to evaluate whether observed PII and PHI transmissions occur within the expected scope of apps. Table~\ref{tab:leakagematrix} illustrates the results of this evaluation by aggregating observed transmissions by their apps' feature category and highlighting transmissions of data types that are not within the feature categories' scopes.

We find that the majority of examined mHealth apps (95.4\%) limit their data collection to the expected specificity. Notable exceptions, however, include two educational apps that collect body measurements such as body weight and height, and five fitness trackers that transmit sensitive medical PHI. Specifically, two fitness trackers collect medical conditions, whereas three others collect the user's heart rate or blood pressure.

A more nuanced evaluation of the observed transmissions of location data reveals that the majority of apps that access and transmit location data do not require this information for their primary features. Of the 14 feature categories we identified, seven explicitly need to locate users to function, such as \textit{Cardio trackers} that record workout routes. Of the 27 apps that transmitted location data, however, 17 do not fall into one of these categories. This includes one \textit{Female health} app that has access to precise location data, enabling it to potentially track the exact location of its users whenever the app is running.

\begin{table}
  \centering
  \caption{Observed transmissions of PII and PHI across distinct app feature categories. Each cell denotes the number of apps of the given feature category that transmitted the data type. \textcolor{blue!7}{\rule{2mm}{2mm}}: data type is within scope of feature category. \textcolor{orange!20}{\rule{2mm}{2mm}}: out of scope, not transmitted. \textcolor{red!30}{\rule{2mm}{2mm}}: out of scope, transmitted.}
  \label{tab:leakagematrix}
  \begin{tabular}{lc|ccc|cc|cc}
    \toprule
    Feature Category & Apps & \rotatebox{90}{Device IDs} &  \rotatebox{90}{Location} &  \rotatebox{90}{User Info} &  \rotatebox{90}{Body measurements} &  \rotatebox{90}{Fitness info} &  \rotatebox{90}{Female health info} &  \rotatebox{90}{Medical info} \\
    \midrule
    Screen Overlay   &  3 & \cellcolor{blue!7}  2 & \cellcolor{red!30}\textbf{1} & \cellcolor{blue!7}2 & \cellcolor{orange!20}0 & \cellcolor{orange!20}0 & \cellcolor{orange!20}0 & \cellcolor{orange!20}0 \\
    Health Education & 14 & \cellcolor{blue!7}  9 & \cellcolor{red!30}\textbf{1} & \cellcolor{blue!7}7 & \cellcolor{red!30}\textbf{2} & \cellcolor{orange!20}0 & \cellcolor{orange!20}0 & \cellcolor{orange!20}0 \\
    Step Counter     &  4 & \cellcolor{blue!7}  4 & \cellcolor{blue!7}1 & \cellcolor{blue!7}4 & \cellcolor{blue!7}1 & \cellcolor{blue!7}1 & \cellcolor{orange!20}0 & \cellcolor{red!30}\textbf{1} \\
    Workout Tracker  & 26 & \cellcolor{blue!7} 25 & \cellcolor{blue!7}1 & \cellcolor{blue!7}25 & \cellcolor{blue!7}7 & \cellcolor{blue!7}10 & \cellcolor{orange!20}0 & \cellcolor{red!30}\textbf{1} \\
    Cardio Tracker   & 14 & \cellcolor{blue!7} 12 & \cellcolor{blue!7}3 & \cellcolor{blue!7}12 & \cellcolor{blue!7}1 & \cellcolor{blue!7}1 & \cellcolor{orange!20}0 & \cellcolor{orange!20}0 \\
    Diet Tracker     & 12 & \cellcolor{blue!7} 12 & \cellcolor{red!30}\textbf{5} & \cellcolor{blue!7}10 & \cellcolor{blue!7}6 & \cellcolor{blue!7}5 & \cellcolor{orange!20}0 & \cellcolor{red!30}\textbf{3} \\
    Wearable         &  3 & \cellcolor{blue!7}  2 & \cellcolor{blue!7}0 & \cellcolor{blue!7}1 & \cellcolor{blue!7}1 & \cellcolor{blue!7}1 & \cellcolor{orange!20}0 & \cellcolor{red!30}\textbf{1} \\
    Mental Wellbeing &  6 & \cellcolor{blue!7}  6 & \cellcolor{orange!20}0 & \cellcolor{blue!7}5 & \cellcolor{blue!7}1 & \cellcolor{blue!7}5 & \cellcolor{orange!20}0 & \cellcolor{orange!20}0 \\
    Pharmacy         &  5 & \cellcolor{blue!7}  5 & \cellcolor{blue!7}2 & \cellcolor{blue!7}4 & \cellcolor{blue!7}1 & \cellcolor{blue!7}0 & \cellcolor{blue!7}0 & \cellcolor{blue!7}0 \\
    Physician Finder &  1 & \cellcolor{blue!7}  1 & \cellcolor{blue!7}1 & \cellcolor{blue!7}1 & \cellcolor{blue!7}0 & \cellcolor{blue!7}0 & \cellcolor{blue!7}0 & \cellcolor{blue!7}0 \\
    Female Health    & 26 & \cellcolor{blue!7} 20 & \cellcolor{red!30}\textbf{5} & \cellcolor{blue!7}19 & \cellcolor{blue!7}4 & \cellcolor{blue!7}1 & \cellcolor{blue!7}9 & \cellcolor{blue!7}7 \\
    Diagnostic       &  1 & \cellcolor{blue!7}  1 & \cellcolor{red!30}\textbf{1} & \cellcolor{blue!7}1 & \cellcolor{blue!7}0 & \cellcolor{blue!7}0 & \cellcolor{blue!7}0 & \cellcolor{blue!7}0 \\
    Health Monitor   & 34 & \cellcolor{blue!7} 23 & \cellcolor{red!30}\textbf{4} & \cellcolor{blue!7}23 & \cellcolor{blue!7}2 & \cellcolor{blue!7}3 & \cellcolor{blue!7}2 & \cellcolor{blue!7}9 \\
    Telemedicine     &  3 & \cellcolor{blue!7}3 & \cellcolor{blue!7}2 & \cellcolor{blue!7}2 & \cellcolor{blue!7}1 & \cellcolor{blue!7}1 & \cellcolor{blue!7}0 & \cellcolor{blue!7}0 \\
  \midrule
  \multicolumn{1}{l}{PHI Specificity} & \multicolumn{1}{l}{} & \multicolumn{3}{c}{Standard} & \multicolumn{2}{c}{NS} & \multicolumn{2}{c}{Medical} \\
  \bottomrule
\end{tabular}
\end{table}

\subsection{Observation vs. Declaration}
\label{subsec:res:declaration}

\begin{figure*}[tb]
  \centering
  \includegraphics[width=\textwidth]{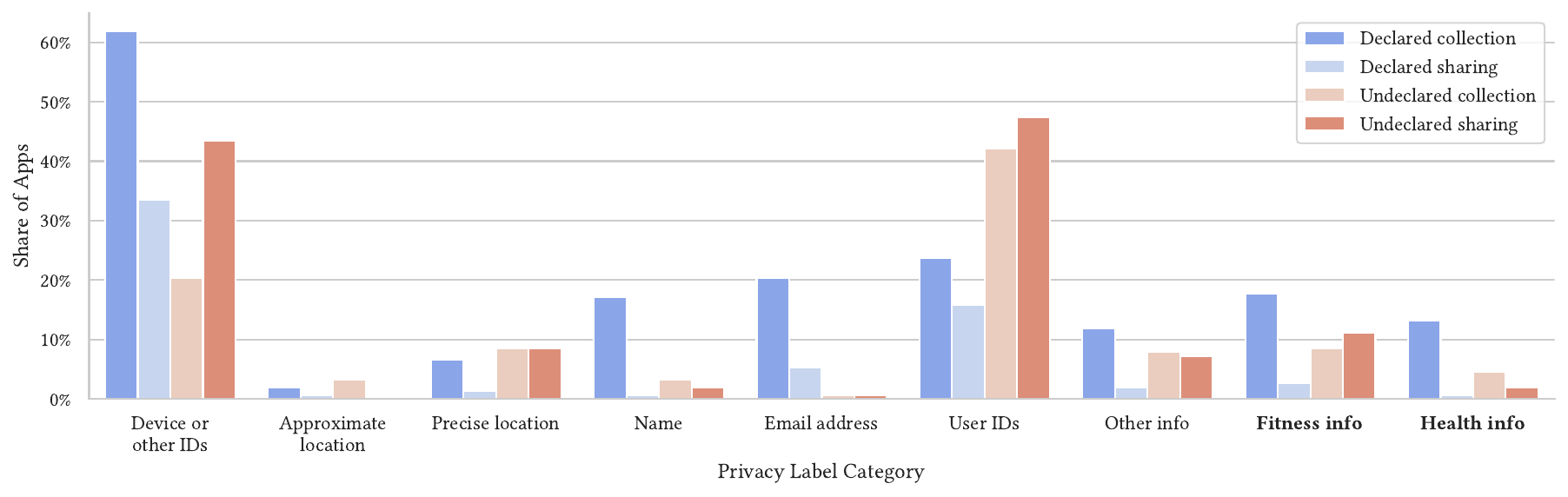}
  \caption{Accuracy of Google Play privacy labels in reflecting mHealth data practices.}
  \label{fig:observation_vs_declaration}
\end{figure*}

Finally, we turn to the comparison of declared and observed data practices by evaluating the accuracy of the privacy labels of the examined mHealth apps.

To this end, we map all types of PII and PHI observed during the examination of the apps' network traffic to the categorisation scheme Google defines for Android privacy labels. This scheme comprises 14 broad categories of PII, of which four are relevant to this study: \textit{Device and other IDs}, \textit{Location}, and \textit{Personal info} cover the types of standard PII we examine, whereas the entirety of PHI falls under \textit{Health and Fitness}. Google further divides these broad categories, defining two subcategories of PHI: \textit{Fitness info} and \textit{Health info}, which are equivalent to nonstandard PHI and medical PHI, respectively. This equivalence simplifies the mapping of observed data types to privacy labels since it enables us to reuse the previously established mapping of data types to PHI specificity.

Android privacy labels distinguish between \textit{data collection} and \textit{data sharing}. In this context, data collection generally refers to the transfer of data off the mobile device, whereas data sharing refers to data transfers to third parties. We build upon this definition, classifying data as \textit{collected} if we observe its transmission to any destination and \textit{shared} if we observe its transmission to a third-party tracker.

Applying this classification to identify declared and undeclared instances of data collection and sharing reveals a significant disparity between declared and actual data practices. Figure~\ref{fig:observation_vs_declaration} illustrates this disparity by contrasting the number of apps that correctly declare their collection and sharing of a given data type with the number of apps that do not declare their data practices accurately.

We identify cases of undeclared collection and sharing for every examined type of PII and PHI. User identifiers emerge as the data type with the worst declaration ratio, with only 25\% of transmissions to third-party trackers declared correctly. Similarly, the majority of apps that share device identifiers and location data do not declare their practices (56.4\% and 81.2\%, respectively).

Despite the added sensitivity of PHI, we observe a high ratio of undeclared transmissions of nonstandard and medical PHI alike. 81\% of apps that share nonstandard PHI, i.e. fitness information, do not publish a corresponding privacy label. Although we only observe a small number of medical PHI transmissions to third-party trackers, 75\% of these are equally undeclared.

Overall, our analysis shows that the examined mHealth apps declare data collection more commonly than data sharing. This is compounded by the fact that privacy labels are often used inconsistently: 15.1\% of apps do not publish any privacy labels, another 18.4\% declare the sharing of some data types without explicitly stating that they collect them, and over 72\% of apps do not correctly declare the full scope of their data practices revealed by our analysis.

\section{Discussion}
\label{sec:disc}

The application of our proposed qualitative analysis framework reveals insights into the privacy practices of the examined mHealth applications that a purely quantitative analysis of their data transmissions alone does not provide. Our findings highlight that third-party tracking is widespread in popular mHealth apps, with over 90\% of apps embedding at least one third-party tracker. In almost all cases, these third-party trackers collect various device and user identifiers that enable them to track users across multiple apps. This becomes a concern in the domain of mHealth, as the fact that individuals use certain apps can reveal sensitive information about them, e.g. the use of a diabetes support app strongly suggests that the user has diabetes.

Although our analysis shows that the majority of examined mHealth apps adheres to the principles of data minimisation and purpose limitation, we still identify instances where apps transmit PHI of a higher specificity than required for the primary features. This observation is particularly concerning in light of the widespread inaccuracy of privacy labels, as it highlights the potential for highly sensitive health information to be leaked to third parties without properly informed user consent.

In general, we find that, although mHealth apps have access to highly sensitive health information not commonly available to other categories of mobile apps, several apps do not fully adhere to the central privacy principles of data minimisation, purpose limitation, and transparency.

Nevertheless, our approach has some limitations. Firstly, despite our best efforts to fine-tune our detection patterns, the keyword-based identification of PII and PHI transmissions in the network traffic of examined apps is unlikely to detect all instances where personal data is transmitted.

Secondly, the derivation of expected data practices, and by extension the evaluation of observed data transmissions through this lens, is limited by the low granularity of the PHI classification provided by the employed app archetype framework.

Thirdly, our evaluation of the transparency of mHealth privacy practices does not consider all avenues used by mHealth apps to communicate their practices to users. Alternatives to privacy labels include formal privacy policies and in-app privacy notices. A more thorough evaluation that considers these alternatives is warranted based on the overall inaccuracy of privacy labels that we observed.

\section{Conclusion}
\label{sec:conc}

Our study introduces a novel qualitative analysis framework that provides a multi-dimensional view of mHealth app privacy practices. This framework evaluates actual data management practices against user expectations and declared privacy policies, identifying significant discrepancies. By analysing 152 leading Android mHealth apps with the proposed framework, we uncover widespread issues, such as health data leakage to third-party trackers and a common disregard for privacy-by-design and transparency principles, despite strict regulations such as the GDPR and HIPAA. Thus, our findings emphasise the need for stricter enforcement of data protection regulations and more accurate privacy labels to inform users properly.

For future research, we aim to expand upon the proposed framework by integrating more sophisticated analytical techniques, such as machine learning models, to improve the automated detection of privacy issues and potential data leakage within mHealth apps. This advancement could significantly improve the scalability of privacy practice assessments, allowing for a broader analysis across the rapidly growing mHealth app market.





\bibliographystyle{plain}
\bibliography{references}

\end{document}